# Severe Plastic Deformation of Ceramics by High-Pressure Torsion: Review of Principles and Applications

Kaveh Edalati,[1,2,3] Jacqueline Hidalgo-Jiménez,[1,2] Thanh Tam Nguyen,[1,3] Hadi Sena,[4] Nariman Enikeev,[5,6,15] Gerda Rogl,[7] Valery I. Levitas,[8] Zenji Horita,[9,10,11] Michael J. Zehetbauer,[12] Ruslan Z. Valiev,[5,6] and Terence G. Langdon[13,14]

[1]WPI International Institute for Carbon Neutral Energy Research (WPI-I2CNER), Kyushu University, Fukuoka, Japan; email: kaveh.edalati@kyudai.jp

[2]Department of Automotive Science, Graduate School of Integrated Frontier Sciences, Kyushu University, Fukuoka, Japan

[3]Mitsui Chemicals, Inc. - Carbon Neutral Research Center (MCI-CNRC), Kyushu University, Fukuoka, Japan

[4]Department of Electrical Engineering, Stanford University, Stanford, California, USA

[5]Ufa University of Science and Technology, Ufa, Russia

[6]Laboratory for Dynamics and Extreme Characteristics of Advanced Nanostructured Materials, Saint Petersburg State University, St. Petersburg, Russia

[7]Department of Materials Chemistry, Faculty of Chemistry, University of Vienna, Wien, Austria

[8]Departments of Aerospace Engineering and Mechanical Engineering, Iowa State University, Ames, Iowa, USA

[9]Graduate School of Engineering, Kyushu Institute of Technology, Kitakyushu, Japan

[10]Magnesium Research Center, Kumamoto University, Kumamoto, Japan

[11]Synchrotron Light Application Center, Saga University, Saga, Japan

[12]Physics of Nanostructured Materials, Faculty of Physics, University of Vienna, Wien, Austria

[13]Materials Research Group, Department of Mechanical Engineering, University of Southampton, Southampton, United Kingdom

[14]Departments of Aerospace and Mechanical Engineering and Materials Science, University of Southern California, Los Angeles, California, USA

[15]Saint Petersburg State Marine Technical University, St. Petersburg, Russia

### Abstract

Ceramics are typically brittle at ambient conditions due to their covalent or ionic bonding and limited dislocation activities. While plasticity, and occasionally superplasticity, can be achieved in ceramics at high temperatures through thermally activated phenomena, creep, and grain boundary sliding, their deformation at ambient temperature and pressure remains challenging. Processing under high pressure via the high-pressure torsion (HPT) method offers new pathways for severe plastic deformation (SPD) of ceramics. This article reviews recent advances in HPT processing of ceramics, focusing primarily on traditional ceramics (e.g., oxides, carbides, nitrides, oxynitrides) and to a lesser extent advanced ceramics (e.g., silicon, carbon, perovskites, clathrates). Key structural and microstructural features of SPD-processed ceramics are discussed, including phase transformations and the generation of nanograins and defects such as vacancies and dislocations. The properties and applications of these deformed ceramics are summarized, including powder consolidation, photoluminescence, bandgap narrowing, photovoltaics, photocatalysis (dye degradation, plastic waste degradation, antibiotic degradation, hydrogen production, $CO_2$ conversion), electrocatalysis, thermoelectric performance, dielectric performance, and ion conductivity for Li-ion batteries. Additionally, the article highlights the role of HPT in synthesizing novel materials, such as high-entropy ceramics (particularly high-entropy oxides), black oxides, and high-pressure polymorphs, which hold promise for energy and environmental applications.

## 1. INTRODUCTION

The topic of severe plastic deformation (SPD), which involves applying very large strains to a material to control its microstructure and enhance its properties, is gaining popularity in the materials science and engineering literature. A recent analysis of publications on this subject indicates that, although the term "severe plastic deformation" occasionally appeared in the titles of some classic papers within the twentieth century, the first international publications with a focus on SPD began only in the 1990s and early 2000s (1). Since 2000, there has been a notable increase in the number of such articles, with the total number of publications exceeding 10,000 in recent years. This surge in interest was driven not only by the promising properties of SPD-processed ultrafine-grained (UFG) or nanostructured materials but also by two other factors. The first was the publication of the first review paper on bulk nanostructured materials in the journal *Progress in Materials Science* in 2000 (2), which was cited over 8,000 times by 2024 according to Google Scholar. Second was the organization of the International Conference on Nanomaterials from SPD (NanoSPD), which was first held in Moscow in 1999 and has been regularly hosted worldwide (the eighth and most recent conference was held in Bangalore, India, in March 2023).

SPD differs from traditional material processing methods such as extrusion, rolling, and drawing because it enables the nanostructuring of materials to form UFG structures with grain sizes below 1 μm. Additionally, SPD allows the formation of various nanostructural features, including defects (e.g., vacancies, dislocations, nanotwins), nanoscale precipitates, and the segregation of alloying elements at grain boundaries. Consequently, nanomaterials processed via SPD may exhibit enhanced multifunctional properties by combining contradictory properties, such as high strength and ductility or high strength and electrical conductivity (3). Furthermore, the range of

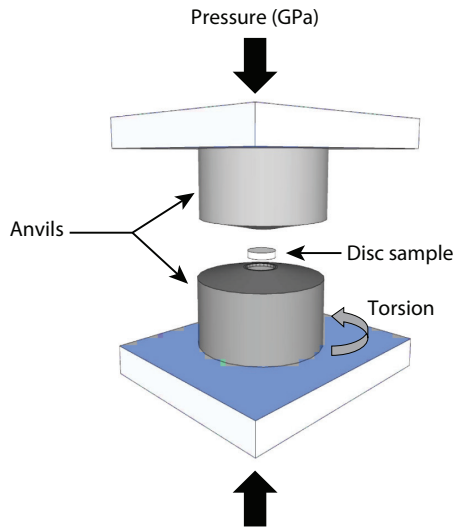

**Figure 1**

Principles of the high-pressure torsion method for severe plastic deformation processing of hard and brittle ceramics.

materials subjected to SPD processing has expanded significantly. The initial focus was on a few pure metals and alloys, followed by intermetallics, polymers, and, more recently, ceramics. Concurrently, the range of SPD techniques has also expanded considerably, with several tens of SPD techniques currently in existence (3, 4). Nevertheless, two techniques employed in early studies, namely equal-channel angular pressing and high-pressure torsion (HPT), remain the most popular.

The processing of low-plasticity and brittle materials, such as ceramics, is of particular interest when employing HPT treatment. In this technique, schematically shown in **Figure 1**, the application of high hydrostatic pressure and the special shape of samples in the form of thin discs prevent fracture, ensuring the maintenance of sample integrity even under significant deformation (5). The high potential of HPT for SPD processing of ceramics was partially highlighted in recent review papers (6, 7), but no review papers have focused exclusively on this issue. This article thus reviews the current research trend of using HPT processing for nanostructure control in ceramic materials, examining the variations in the materials' phase transformations, microstructures, and resultant properties and applications.

## 2. PLASTIC DEFORMATION OF CERAMICS

Ceramics exhibit distinct deformation mechanisms at room and high temperatures, primarily due to their atomic structure and bonding. At room temperature, ceramics are extremely brittle because of their strong ionic and covalent bonds, which make dislocation motion (i.e., the primary mode of deformation in metals) very difficult. As a result, when ceramics are subjected to stress, they tend to fracture rather than deform plastically. There has been limited success in enhancing the dislocation nucleation and activity in ceramics, although a recent study suggested a strategy to overcome the inherent brittleness of ceramics in which dislocations are generated by an external source, such as interfaces with a metallic phase (8). Besides limited dislocation activity, microcracks, which can be inherent from the manufacturing process or induced by stress, propagate quickly, leading to sudden and brittle failure of ceramics. Grain boundaries in ceramics do not effectively hinder crack propagation, further contributing to their brittleness. At high temperatures, the increased atomic mobility can allow some ceramics to exhibit plastic deformation, but this typically occurs only at temperatures close to their melting point. This issue was investigated by creep experts, who shed light on the deformation mechanisms and established the deformation maps of ceramics. Creep is the nonrecoverable strain occurring in a crystalline solid when it is subjected to an imposed load over a prolonged period of time. Measurements of creep flow in metals date back to 1910 (9), and many experiments showed that in creep there is an initial primary stage where the strain rate decreases, a secondary or steady-state stage where the rate remains reasonably constant, and then a tertiary stage where the rate increases to fracture. Within the steady-state region in metals, there is usually power-law creep, where the rate varies with the applied stress raised to a power $n$ that is typically close to $\sim 5$ due to the glide and climb of dislocations. For polycrystalline metals at very low stresses, there are excess vacancies along boundaries perpendicular to the tensile axis, the vacancies are depleted along boundaries experiencing a compressive stress, and this produces vacancy flow either through the crystalline lattice in Nabarro–Herring creep (10, 11) or along the grain boundaries in Coble creep (12), with strain rates varying linearly with stress so that $n = 1$.

Ceramics were generally considered inherently brittle, and investigations of their creep behavior started much later, but several early results in compression or bending established that Nabarro–Herring creep was dominant in $Al_2O_3$ (13), BeO (14, 15), MgO (16), and $UO_2$ (17, 18), although there was scattered evidence for a power-law creep at the higher stresses in $UO_2$ (19, 20) and $ThO_2$ (20). The importance of power-law creep was subsequently confirmed by conducting compressive creep experiments on polycrystalline LiF (21) and by testing LiF samples having grain sizes differing by more than an order of magnitude (22). Other experiments confirmed power-law creep in MgO (23), NaCl–KCl solid-solution alloys (24), and $UO_2$ (25). Measurements of grain shape changes during creep suggested that all strain arose from grain boundary sliding in MgO (26) and a U–Pu carbide (27), but these results were erroneous because migration tends to spheroidize the grains during creep (28). Later experiments gave reasonable estimates for the sliding contributions by measuring the vertical offsets between adjacent surface grains with either interferometry (29) or a surface analyzer (30). By using the conventional power-law creep equation, it was also possible to estimate the stacking fault energies of ceramic materials from creep data (31). A complexity in the creep of ceramics is the occurrence of ambipolar diffusion where conventional lattice and grain boundary diffusion creep are controlled by the movement of either the anions or the cations, and these ionic species may diffuse along different paths within the solid (32). This approach was developed for creep of MgO and $Al_2O_3$ (33), and the diffusion creep processes can be depicted in a simple way by constructing deformation mechanism maps of logarithmic grain size against the reciprocal of temperature, as in **Figure 2** for (*a*) $Al_2O_3$ and (*b*) MgO (34). Using creep data available for ceramics, reviews have been prepared covering their mechanical characteristics (35) and flow mechanisms (36).

Polycrystalline materials tested in tension usually break at relatively low strains, but under some conditions, they may pull out to strains of more than 400%, which is defined as superplastic flow (37). The first reports of superplasticity in ceramics were for polycrystalline yttria-stabilized tetragonal $ZrO_2$ (Y-TZP) (38) and a Y-TZP/$Al_2O_3$ composite (39) where the elongations were >120% and >200%, respectively. These elongations are high, but they are not superplastic because they fail to reach at least 400%. A later report documented a true superplastic elongation of 800% in Y-TZP, as shown in **Figure 2c** (40), and there are now reports of high superplastic elongations in ceramics, including of 2,510% in a $ZrO_2$–$Al_2O_3$–spinel composite (41).

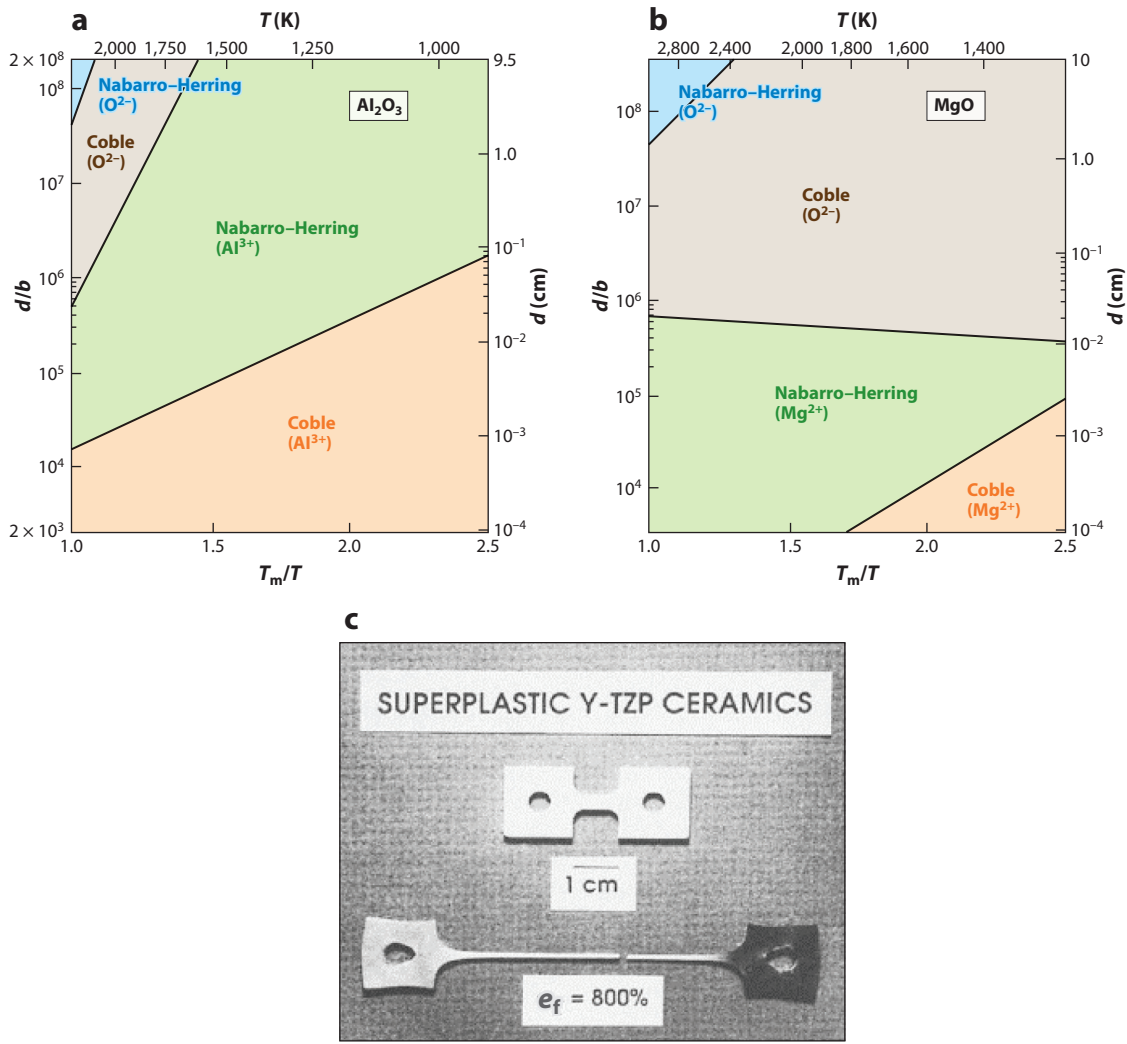

**Figure 2**

(*a*,*b*) Deformation mechanism maps showing the effect of ambipolar diffusion in (*a*) $Al_2O_3$ and (*b*) MgO. (*c*) Superplasticity in yttria-stabilized tetragonal $ZrO_2$ (Y-TZP) where $e_f$ indicates elongation to failure. Panels *a* and *b* adapted with permission from Reference 34; copyright 1980 Elsevier. Panel *c* adapted with permission from Reference 40; copyright 1990 Elsevier.

## 3. SEVERE PLASTIC DEFORMATION OF CERAMICS

Efforts to severely deform ceramics have a long history, especially due to the relevance of ceramics in geological studies. A significant milestone was reached in 1935 when Bridgman (42) identified high pressure in the gigapascal range as crucial for achieving plastic deformation in ceramics, noting that pressures induce plasticity and significantly suppress crack propagation. Bridgman expanded Boker's 1915 work (43) by combining high pressure with torsional shear to develop the first HPT facility for SPD processing of ceramics (42). In HPT, high pressure helps to suppress crack formation and propagation, while the shear stress helps to drive dislocation motion to achieve plasticity (44). Nowadays, it is known that the plastic deformation of ceramics under high pressure at ambient temperature involves a combination of mechanisms such as dislocation activity,

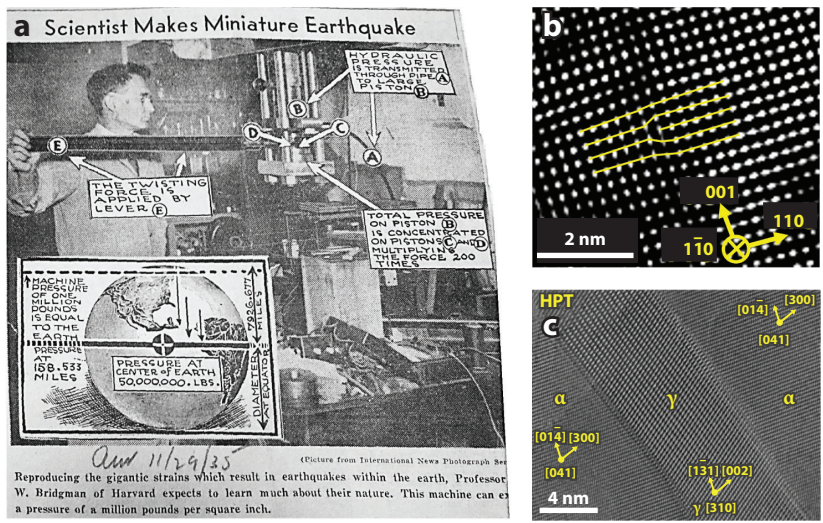

**Figure 3**

(*a*) The first press release by Bridgman about severe plastic deformation processing of ceramics by high-pressure torsion (HPT) highlighted the occurrence of earthquake-like shocks due to the microcracking and healing deformation mechanism. (*b,c*) Lattice images indicating deformation mechanisms during HPT processing based on (*b*) dislocation activity in $ZrO_2$ and (*c*) phase transformation in $Al_2O_3$. Panel *a* reproduced with permission from Reference 44; copyright 2015 Elsevier. Panel *b* adapted with permission from Reference 47; copyright 2011 Acta Materialia. Panel *c* adapted with permission from Reference 48; copyright 2018 Acta Materialia.

phase transformation, amorphization, microcracking and healing, and twin formation, while grain boundary sliding and diffusional creep can occur with increasing temperature. The high-pressure environment enables mechanisms that are otherwise suppressed in ceramics, allowing them to accommodate plastic strain and deform more like ductile metals (45, 46). The mechanism of the microcracking and healing of ceramics leads to snapping and noisy shocks during plastic deformation, which Bridgman suggested could mimic deep-seated earthquake mechanisms—a notion that was highlighted in the press under the title "Scientist Makes Miniature Earthquake," as shown in **Figure 3*a*** (44). **Figure 3*b,c*** shows two micrographs depicting the deformation of ceramics by dislocations (47) and phase transformation (48), respectively.

The advent of HPT spurred further research on ceramics throughout the last century both by Bridgman and colleagues (49–51) and other scientists in the fields of geology and physics (52–59). In 1960, Griggs (52) contributed to the fabrication of an HPT machine with similar operation and features to current facilities. Researchers such as Bates et al. (53), Bell (54), Dachille & Roy (55), Miller et al. (56), and Vereshchagin et al. (57) reported that SPD under pressure could accelerate phase transformations. Morozova et al. (58, 59) found that high-pressure shearing could reduce oxides, a finding initially observed by Bridgman (42, 49), indicating the formation of large fractions of oxygen vacancies. The integration of HPT with diamond anvil cells (DACs) and the subsequent development of shear (rotational) DACs (RDACs) in the 1980s (60, 61) significantly advanced this research field by enabling higher processing pressures and in situ examinations (62–64). It should be noted that although high-energy ball milling, introduced by Benjamin in the 1970s (65), can introduce deformation combined with alloying, fragmentation, and welding, achieving pure plastic deformation in ceramics with this method is challenging due to limited applied pressures (66, 67). Despite considerable efforts by geologists and physicists using the HPT method, SPD processing of ceramics was largely overlooked by materials scientists until a 2010 publication about alumina increased interest in using SPD to achieve functional properties in ceramics (68). Nowadays, SPD is applied not only to traditional ceramics such as oxides, nitrides, and carbides but also to various advanced ceramics, including silicon semiconductors (62, 69), carbon polymorphs (62, 63), and hydrides (70, 71).

## 4. PHASE TRANSFORMATION IN SEVERELY DEFORMED CERAMICS

The study of phase transformation in severely deformed materials has been well stimulated owing to the HPT process. A major advantage of the HPT process is that it not only introduces intense plastic strain in materials but also imposes high pressure. The materials may undergo phase transformation if they take different crystal structures with a change in pressure and/or strain. Furthermore, the high-pressure phases may be retained at ambient pressure even after the applied pressure is released, or the phase transformation may be facilitated due to the presence of intense strain. In this section, some results reported for phase transformations of ceramics by ex situ observations after HPT processing and in situ observations using RDAC are reviewed.

### 4.1. Ex Situ Observations of Phase Transformations After High-Pressure Torsion

Bridgman, a pioneer in combining high pressure with shear strain (42), investigated the effects of pressure on crystal structures in a variety of materials, as discussed in a comprehensive review paper about Bridgman's research activities (44). Bridgman's main concern was phase transformations not only in metallic materials but also in many organic and inorganic materials, including ceramics. Recent advances in phase transformations of ceramics by HPT processing were partly documented in References 6, 7, 46, and 72, including where $ZrO_2$ and $Al_2O_3$ transform from metastable to stable phases and $BaTiO_3$, $ZnO$, $TiO_2$, $Y_2O_3$, $SiO_2$, and $VO_2$ transform from stable to metastable phases, for example, as shown in **Figure 4a** for $Y_2O_3$ (73). The former case occurs because the activation barrier for the transformation is lowered by intense shear strain, whereas the latter case occurs due to the creation of stress–strain fields favorable for the high-pressure phases. High

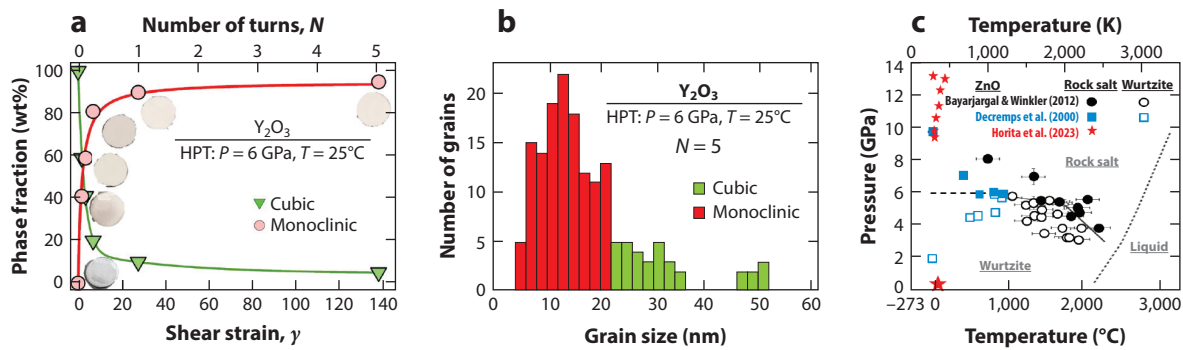

**Figure 4**

(*a*) Influence of shear strain on the fraction of the high-pressure monoclinic phase and (*b*) influence of grain size on the stability of the high-pressure monoclinic phase of $Y_2O_3$ after processing by high-pressure torsion (HPT) and releasing pressure. (*c*) Temperature–pressure phase diagram of $ZnO$, where outlined and filled black circles are results from Bayarjargal & Winkler (78) for wurtzite and rock-salt structures, respectively; outlined and filled blue squares are from Decremps et al. (79), representing transformation points from rock salt to a wurtzite structure and from wurtzite to a rock-salt structure, respectively; and star symbols represent the results of Horita et al. (77), showing that the rock-salt structure reaches 100% of the materials. Panels *a* and *b* adapted with permission from Reference 73; copyright 2017 American Chemical Society. Panel *c* adapted from Reference 77.

densities of dislocations and/or grain boundaries are major causes of such a stress–strain field, and it appears that there is a critical grain size below which high-pressure phases can exist at ambient pressure, as shown in **Figure 4*b*** (73).

Besides the fundamental significance of phase transformation by SPD processing, special attention has been paid to enhancing properties associated with phase transformations (6, 7, 46). For example, the dielectric property of $BaTiO_3$ is significantly increased by its transformation from a tetragonal to cubic structure (74), as are the photocatalytic properties of $TiO_2$ by a transformation to a columbite structure (75) and of ZnO by a transformation to a rock-salt structure (76). An in situ high-energy X-ray diffraction (XRD) experiment using a synchrotron facility demonstrated that ZnO with a 100% fraction of rock-salt structure was produced even at ambient pressure when HPT-processed samples were subjected to high pressures at temperatures of ~200°C. The results based on this in situ XRD analysis are plotted in the pressure–temperature phase diagram shown in **Figure 4*c*** (77–79). It was then anticipated that ZnO could be more efficiently used as a photocatalyst under visible light. To attain enhanced photocatalytic properties, the HPT process was further used to synthesize mixtures of $TiO_2$–ZnO (80) and of GaN–ZnO (81). HPT was also conducted as a mechanical alloying process with subsequent oxidation for achieving a $TiZrHfNbTaO_{11}$ photocatalyst (82, 83) and with oxidation plus nitriding for synthesizing a $TiZrHfNbTaO_6N_3$ photocatalyst (84).

It should be noted that the HPT process can chemically reduce the oxides by introducing a large fraction of oxygen vacancies so that the bandgap may be controlled to enhance the photocatalytic effect such as in $BiVO_4$ (85). Bridgman (49, 86) also reported the reduction of $Bi_2O_3$ to metallic bismuth, $SnO_2$ to SnO, and $MnO_2$ to MnO because of a change in valence through intense shear as if the chemical reaction took place. Recently, it was shown that the HPT process is useful in creating a Magneli structure with $Ti_nO_{2n-1}$ ($n = 3$–10) (87). It was demonstrated that the HPT process can control the valence of titanium in a mixture of alumina and titania powders and, thus, lead to an increase in the superconducting transition temperature (87, 88).

## 4.2. In Situ Observations of Phase Transformations in Rotational Diamond Anvil Cells

SPD-induced phase transformations can occur in DACs and RDACs and can be investigated by in situ observations (63, 89, 90). This concept and the corresponding three-scale theory were introduced in Reference 91 and elaborated in detail in References 44, 92, and 93. Unlike traditional pressure-induced phase transformations, strain-induced phase transformations occur at defects generated during plastic flow, most likely at the tip of dislocation pileups, producing the strongest stress concentrator, as shown in **Figure 5*a*** (91, 94). They require completely different thermodynamic and kinetic description (62, 91) and experimental characterization (95–98) than conventional stress- and pressure-induced phase transformations (99–103). In this section, some results examined during SPD in DACs and RDACs using in situ XRD are reviewed.

Under hydrostatic loading, pressure-induced amorphization in hexagonal 6H-SiC was not observed at any pressure; no transformation occurs up to 95 GPa in DACs (104). In RDACs, reversible phase transformation to a new high-density amorphous SiC phase was found at ~30 GPa and rotation of an anvil of 2,160°, as shown in **Figure 5*b***, thus exhibiting a polymorphism.

The highly ordered and textured rhombohedral BN, compressed along the hexagonal *c* axis, transformed to the superhard cubic BN (105, 106) at 5.6 GPa instead of 55 GPa under hydrostatic pressure. Continuum thermodynamic theory (106), utilizing general theory for phase transformations in elastoplastic materials (99), described this new phenomenon of phase transformation induced by rotational plastic instability (106). It was reported that shear in an RDAC does not change the

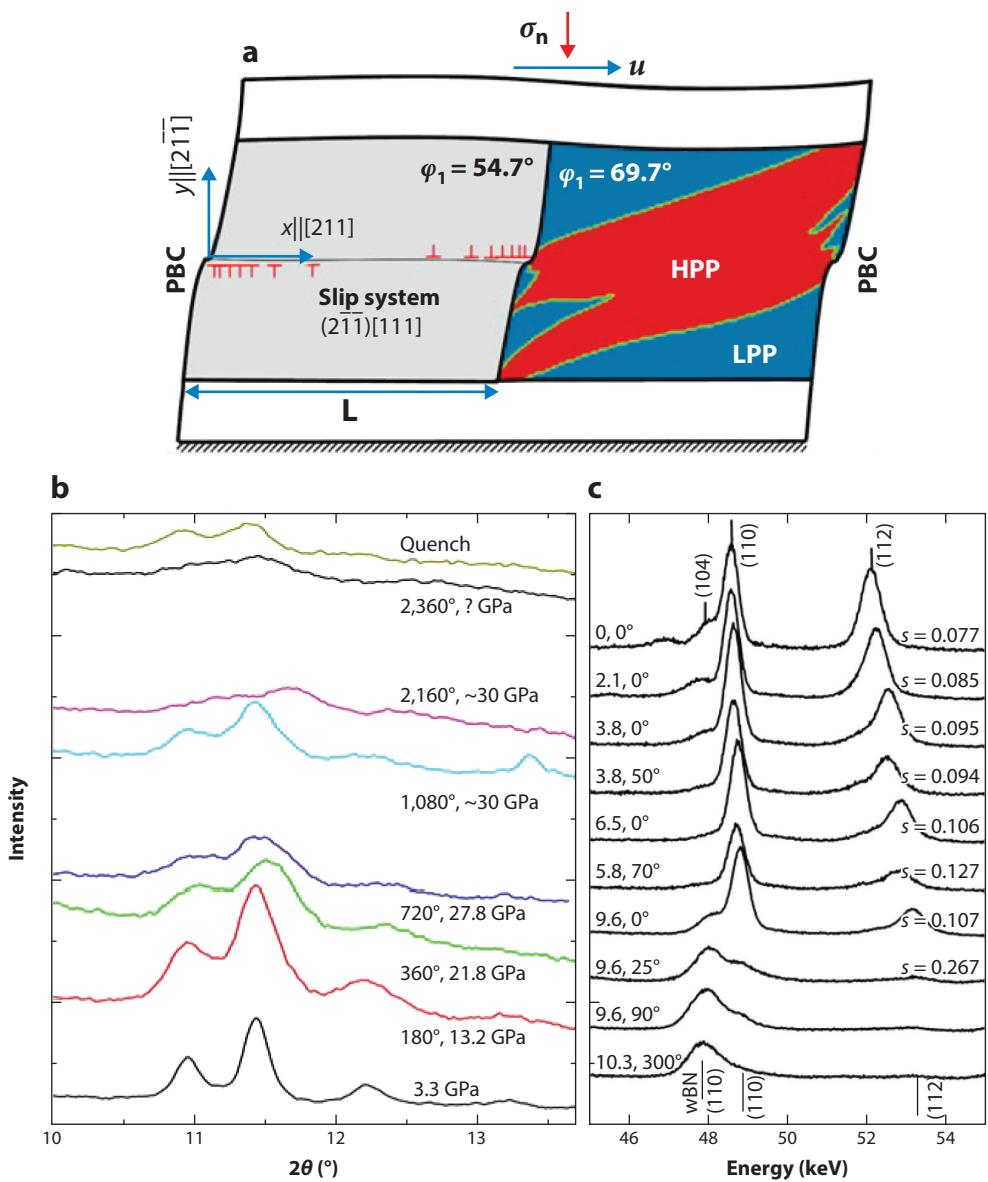

**Figure 5**

(*a*) Phase-field simulations of nucleation and growth of a high-pressure phase (*red*) in a bicrystal at the tip of the dislocation pileup under compression and shear. Panel adapted with permission from Reference 93; copyright 2020 Acta Materialia. (*b*) X-ray diffraction patterns of 6H-SiC at different pressures and rotation angles in rotational diamond anvil cells (shown near curves) and quenching to normal pressure (the question mark means that pressure was not determined). Panel adapted with permission from Reference 104; copyright 2012 American Physical Society. (*c*) Evolution of X-ray diffraction patterns of hexagonal BN with increasing pressure (in GPa) and rotation angle, as well as dimensionless concentration of turbostratic stacking faults *s* (all shown near curves) during hexagonal→wurtzite phase transformation in BN. Panel adapted from Reference 107 with permission from AIP Publishing. Abbreviations: BN, boron nitride; HPP, high-pressure phase; LPP, low-pressure phase; PBC, periodic boundary condition; wBN, wurtzite-type BN.

phase transformation pressure from the hexagonal BN to superhard wurtzitic BN (∼10 GPa) (107). It has been found that a growing concentration of turbostratic stacking faults compensates for the reduction in phase transformation pressure due to SPD, as shown in **Figure 5c**. Phase transformation of BN from highly disordered hexagonal to wurtzite was obtained in an RDAC at 6.7 GPa but was not observed under hydrostatic pressure up to 52.8 GPa (108).

While olivine [$(Mg_{0.91}Fe_{0.09})_2SiO_4$] was never transformed to spinel at room temperature at any pressure, olivine–ringwoodite phase transformation in RDACs was obtained at 15–28 GPa (98). The transformation pressure linearly reduces with increasing plastic strain, dislocation density, and microstrain and decreasing crystallite size. The main rules of the coupled SPD, phase transformation, and microstructure evolution were found to be similar to those for zirconium (95, 96). The obtained results support the new mechanism of deep-focus earthquakes (109).

Results strongly supporting a dislocation pileup–based nucleation mechanism in silicon (44, 91–95) were obtained for strain-induced phase transformations, as shown in **Figure 5a** (109). A correlation between the particle size's direct and inverse Hall–Petch effect on yield strength and pressure for strain-induced phase transformation was found. Pressure in small Si-II and Si-III regions is 5–7 GPa higher than in Si-I. For 100-nm particles, the strain-induced phase transformation Si-I→Si-II initiates at 0.3 GPa versus 16.2 GPa under hydrostatic conditions; Si-I→Si-III transformation starts at 0.6 GPa and does not occur under hydrostatic pressure. Retaining Si-II and single-phase Si-III at ambient pressure and obtaining reverse Si-III→Si-I phase transformation demonstrated the possibilities of manipulating different synthetic paths (110). Note that retaining phases at ambient pressure was attributed to the effect of the nanograin size in $Y_2O_3$ and $TiO_2$ during HPT by ex situ studies (72, 73, 75, 76, 111), while some studies reported a more complicated contribution of different microstructural factors in stabilizing phases in ZnO (72, 77), $Bi_2O_3$ (112), and $TiO_2$ (113), as reviewed elsewhere (46).

In DAC experiments, graphite transforms to cubic diamond at ∼70 GPa under quasihydrostatic compression; in RDACs, it transforms to hexagonal and cubic diamonds at ∼20 GPa (89). A modeling of the phase transformation at the tip of the dislocation pileup demonstrated the possibility of drastically reducing the external pressure, even down to ambient pressure (44, 91, 92). This modeling inspired RDAC experiments at very low pressure, resulting in a reversible phase transformation to the hexagonal diamond at 0.4 GPa and an irreversible phase transformation to the cubic diamond at 0.7 GPa (114). At 3 GPa and by applying shear, a new orthorhombic diamond appeared. After pressure release, the samples contained cubic and orthorhombic diamond phases, fullerenes, amorphous phases, and fragmented graphite. Molecular dynamics simulations for nanocrystalline graphite showed that shear can reduce the external pressure for phase transformation at 950 K down to 2.3 GPa (115). Structural changes in graphene in RDACs were also reported (116).

Two new phases, IV and V, of fullerene $C_{60}$ were obtained in RDACs (62, 117). It was claimed that phase V of fullerene is harder than a natural diamond. Phase transformations in Reference 90 are consistent with those in Reference 117; however, due to the lower pressure used in Reference 90, only phase IV was found, which scratches the diamond anvil surface. Pressure growth in RDACs at constant force despite a significant volume decrease can be explained theoretically (91, 118). Fullerene in RDACs exhibited localized shear bands containing five high-pressure phases (119).

In transformation-induced plasticity (TRIP), it was theoretically predicted (99, 103) that shear strain $\gamma$ can be determined by the equation

$$\gamma = \frac{2}{\sqrt{3}} \, |\varepsilon_0| \left(\tau/\tau_y\right) \sqrt{1 - \left(\tau/\tau_y\right)^2},$$

where $\varepsilon_0$ is the volumetric strain during phase transformation, $\tau$ is the shear stress, and $\tau_y$ is the yield shear strength. In shear bands, shear stress becomes comparable to the yield shear strength (i.e., $\tau \to \tau_y$), causing the shear strain to approach infinity (i.e., $\gamma$ becomes very large). Indeed, TRIP exceeding the traditional plasticity by a factor of 20 was determined for the hexagonal→wurtzite phase transformation in BN (107). The positive mechanochemical feedback between phase transformations, disordering, conventional plasticity, and TRIP led to cascading structural changes. TRIP was utilized to resolve the contradictory appearance of shear-transformation bands in fullerene, notwithstanding the much stronger high-pressure phases (119). Reaction-induced plasticity was revealed during chemical reactions in Ti–Si and Ni–Si mixtures in shear bands (120). TRIP, along with self-blown-up TRIP heating in shear bands, was recently utilized to resolve the main puzzles in the mechanisms of the deep-focus earthquake (109).

The reviewed results have multiple potential applications, including defect-induced synthesis of new nanostructured materials and phases, as detailed in each corresponding paper. Note that more advanced experimental and computational studies used in DACs and RDACs for metals should be repeated for ceramics to check whether similar rules are valid (95–97).

# 5. MICROSTRUCTURAL FEATURES OF SEVERELY DEFORMED CERAMICS

It is well known that SPD can introduce various kinds of point defects such as vacancies (121), linear defects such as dislocations (122), and planar defects such as grain boundaries (123) in metallic materials. Such microstructural changes can be even more pronounced in ceramics after HPT processing at ambient temperature (46). In this section, the evolution of vacancies, dislocations, and nanograins in severely deformed ceramics is discussed.

## 5.1. Vacancies

Vacancies play a significant role in modifying the electronic structure of ceramics. These modifications are useful in a wide range of applications, such as electronics (124), optics (125), storage devices (126), sensors (127), fuel cells (128), and thermoelectrics (129). Metal oxides are among the most widely used ceramics, with oxygen vacancies having a substantial impact on their properties. Typically, oxygen vacancies form localized energy levels above the valence band maximum in semiconductors and insulators. Although oxygen vacancies can be induced by various external stimuli (e.g., high-energy light irradiation, thermal stress, pressure), doping is the most common method that can lead to the formation of oxygen vacancies if the oxidation state of the dopant is higher than that of the host cation (130–133). SPD has been demonstrated to be a new method for inducing vacancies in ceramics.

Oxygen vacancies have been successfully introduced by HPT processing into ceramics such as $TiO_2$ (75, 111, 113, 134, 135), ZnO (76, 136), $Y_2O_3$ (73), $BiVO_4$ (85), $TiO_2$–ZnO (80), $CsTaO_3$ (137), $LiTaO_3$ (137), $Bi_2O_3$ (112), $ZrO_2$ (138), $Al_2O_3$ (139), MgO (140), $VO_2$ (141), $BaTiO_3$ (74), and $SiO_2$ (142). The presence of vacancies in these ceramics was proven using different methods such as XRD, Raman spectroscopy, X-ray photoelectron spectroscopy, electron energy loss spectroscopy, and electron spin resonance analysis. The mechanism of vacancy formation in ceramics by HPT is strain-induced, and, thus, the fraction of oxygen vacancies usually increases when increasing the applied shear strain (i.e., number of HPT turns), as shown in **Figure 6a**. The increase in the fraction of oxygen vacancies significantly reduces the optical bandgap of metal oxides, which is advantageous for applications such as photocatalysis. HPT does not have the limitations of the doping methods for oxygen vacancy generation. One limitation of the doping method is that dopants with high solubility often alter the chemical composition and structure of the host

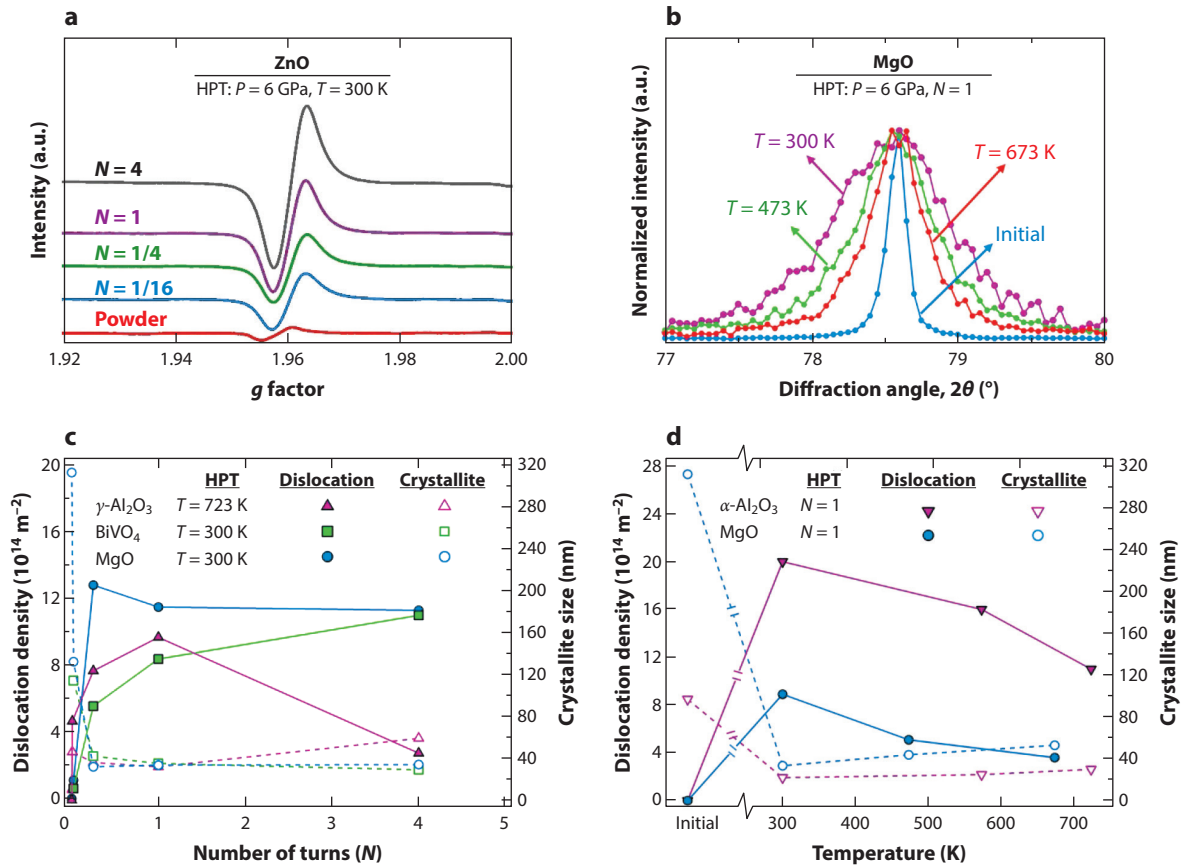

**Figure 6**

(*a*) Electron spin resonance spectra of ZnO before and after high-pressure torsion (HPT) processing for various turns (higher peak intensity indicates higher oxygen vacancy concentration). Data from Reference 136. (*b*) X-ray diffraction (XRD) profiles for the (222) atomic plane of MgO before and after HPT processing for different processing temperatures. Data from Reference 139. (*c,d*) Dislocation density (*solid lines*) and crystallite sizes (*dashed lines*) versus (*c*) number of turns and (*d*) processing temperature for $\alpha$-Al$_2$O$_3$, $\gamma$-Al$_2$O$_3$, BiVO$_4$, and MgO ceramics processed by HPT. Panel adapted from Reference 149 (CC BY 4.0).

material, while dopants with low solubility tend to segregate undesirably. Additionally, high temperatures are typically required for doping, which can result in the unintended precipitation of dopants. Here, it should be noted that the introduction of vacancies through SPD is not limited to oxides. Carbon vacancies in graphite (143) and nitrogen vacancies in the GaN–ZnO oxynitride (81) are other examples.

## 5.2. Dislocations

Ceramics are typically considered brittle under conventional mechanical loading. However, their plasticity, driven by dislocation motion, has been documented for over 60 years through methods such as indentation experiments and the compression of single crystals at elevated temperatures (144). The crystal lattice of ceramics, consisting of packed ions, can be significantly influenced by dislocations, affecting their functional properties (145, 146). Nonetheless, the conditions promoting dislocation activity in most ceramics are rather exceptional. SPD under high pressures greatly expands the possibilities for inducing dislocations in ceramics, thereby allowing modification of their properties (5, 147, 148). Quantitative evaluation of dislocation density in SPD-processed ceramics remains a crucial research challenge, despite electron microscopy techniques confirming the presence of dislocations in highly distorted nanostructures of SPD-processed ceramics (7). Given the lack of quantitative dislocation analyses in the literature, this section presents recent results on this topic (149).

Several XRD approaches can yield reliable quantitative data on dislocation density in polycrystalline materials with different crystal symmetries. While sophisticated methods such as convolutional multiple whole profile fitting (150) and whole powder pattern modeling (151) offer comprehensive XRD analysis for various dislocation configurations, researchers often prefer simpler procedures for quick dislocation density estimation. For materials with cubic lattices, in the simplest case, dislocation density ($\rho$) is calculated using the Burgers vector (**b**), coherent domain size ($d$), and microstrain ($\varepsilon$) (152):

$$\rho = 2\sqrt{3} < \varepsilon^2 >^{1/2}/(\mathbf{b}d). \hspace{2cm} 1.$$

Materials with other symmetries are more complex to analyze due to having multiple Burgers vectors. For instance, in hexagonal close-packed (HCP) lattices, different dislocation families with varying Burgers vector values must be considered. In such cases, approaches such as the one proposed in Reference 153 are useful, providing different expressions for dislocation densities with $a$ and $c$ lattice parameter components. Broadening values can be calculated using the anisotropic Popa line broadening model (154), implemented via Rietveld refinement software such as MAUD (Material Analysis Using Diffraction) (155). This method has successfully estimated dislocation density in SPD-processed HCP alloys (156, 157), though its efficiency for ceramics remains unverified. Given that ceramics often have crystal lattices with even lower symmetries, rigorous XRD analysis is quite complicated. Therefore, researchers use the following equation for a rough estimation of dislocation density in ceramics (152):

$$\rho = 1/d^2. \hspace{2cm} 2.$$

This section analyzes XRD data reported in the literature and additional data provided by the corresponding authors for MgO (140), $Al_2O_3$ (48), and $BiVO_4$ (85) after HPT processing at various turns ($N = 0.25, 1, 4, 5$) and temperatures ($T = 300, 473, 573, 673, 725$ K), using MAUD software (155). Dislocation density was calculated using either Equation 1 or 2. In the latter case, the entire peak broadening was considered to be induced by the size effect. The results are summarized in **Figure 6b–d**. Analysis of peak profiles in **Figure 6b** shows that HPT processing of MgO at ambient temperature leads to a significant increase in peak broadening, while increasing the processing temperature reduces peak broadening, indicating a significant contribution of diffusional processes to dislocation rearrangements in ceramics. Quantitative XRD analysis results in **Figure 6c** as a function of applied strain highlight several points. (*a*) For MgO and $BiVO_4$, crystallite size decreases and quickly saturates, even after just one-fourth of an HPT rotation. It should be noted that crystallite size, referring to the coherent domain size determined by XRD, differs somewhat from grain size, which refers to the distance between grain boundaries measured by microscopy methods. (*b*) Dislocation density in MgO and $BiVO_4$ increases dramatically after HPT processing, reaching an ultrahigh level of about $10^{15}$ m$^{-2}$. Such high densities have been observed in limited ceramics, such as high-entropy oxide ceramics produced by ball milling followed by cold isostatic pressing and annealing (158). This high dislocation density aligns with theoretical predictions of the upper limit of dislocation density achievable in single-crystal ceramics (159). (*c*) γ-$Al_2O_3$ shows a different trend, with an initial increase in dislocation density (and a decrease in crystallite size) followed by a decrease in dislocation density (and an increase in crystallite size) at higher numbers of HPT turns. These changes at a processing temperature of 773 K are attributed to a strain-induced γ→α phase transformation (48). The temperature dependence of structural parameters in **Figure 6d** for MgO and α-Al₂O₃ shows a reasonable trend: Crystallite size increases while dislocation density decreases with rising temperature. These quantitative analyses confirm the high dislocation density in SPD-processed ceramics, with levels comparable to those reported in severely deformed metallic alloys (7). Such high dislocation densities are expected to contribute to enhancing the plasticity of ceramics, as discussed in a recent publication (8).

### 5.3. Nanograins

One of the most common microstructural changes by SPD is the formation of a UFG or nanograined structure, which has been frequently reported in metallic materials (2, 3). This effect was also reported in HPT-processed ceramics such as BaTiO₃ (74), Bi₂O₃ (112), BiVO₄ (85), TiO₂ (75, 135, 160), Y₂O₃ (73), and ZnO (76, 136). **Figure 7a** shows an example of grain refinement to the nanometer in ZnO after HPT processing (136). Although the mechanisms for grain refinement in ceramics during SPD are not clear, it is assumed to be similar to metals (123), involving dislocation generation, dislocation accumulation, dynamic recrystallization, and grain boundary migration (46). In pure metals with metallic atomic bonding processed by HPT, the grain size reaches a steady state at the submicrometer level (the micrometer level for metals with high melting points), and the steady-state grain size is closely related to the melting point ($T_m$). Previous studies showed that the smallest grain size in metals with metallic bonding is achieved in metals with low melting points as a result of temperature-dependent diffusion, dislocation recovery, and recrystallization (161). In ceramics, the steady-state grain size after HPT processing goes down to the nanometer level. **Table 1** shows a summary of grain sizes (distance between grain boundaries measured by transmission electron microscopy) for several ceramics processed by HPT including their atomic bond energy and melting points (47, 48, 80, 81, 111, 134, 137–142, 162–167). In all ceramics, grain size is reduced after HPT processing, with the notable exception of nanograined γ-Al₂O₃ processed by HPT at 773 K, in which grain size increases due to a strain-induced γ→α phase transformation (48).

To clarify the differences in grain refinement between oxides and metals, the steady-state grain size was plotted versus the homologous temperature ($T/T_m$) in **Figure 7b** and versus atomic bond energy in **Figure 7c**. The data for metals were gathered from three publications and references therein (161, 168, 169). For metals, grain sizes decrease with decreasing homologous temperature or increasing atomic bond energy but do not reach the nanometer level. While a systematic trend cannot be found for ceramics, their grain sizes are always at the nanometer level. This should be due to the covalent or ionic nature of atomic bonds in ceramics, which inhibits dislocation activity. Dislocation movement in ceramics with limited slip systems is difficult, and their Peierls barrier is high, so thermal activation is usually required (144, 170). A closer look at the behavior of oxides shows a weak relation between bond energy and grain size, with an unexpectedly larger grain size for higher atomic bond energies. This can be justified by considering that a higher Peierls barrier needs to be overcome for higher atomic bond energies to promote dislocation movement and recrystallization, which are essential for refining grains.

## 6. PROPERTIES AND APPLICATIONS OF SEVERELY DEFORMED CERAMICS

The properties and applications of ceramics are significantly influenced by their phase transformations, defects, and grain size. All these factors can be controlled through SPD processing. Metastable phases often exhibit interesting properties under ambient conditions, such as the high hardness of diamond compared with graphite or the high toughness of the tetragonal phase of

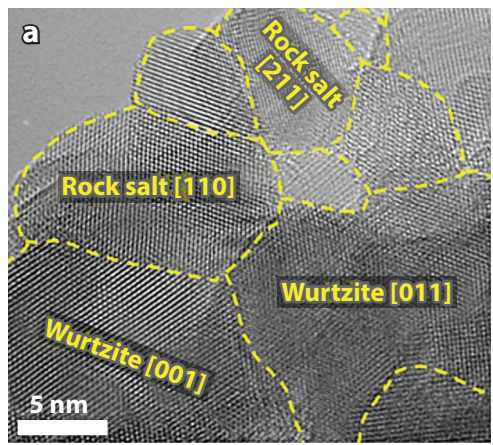

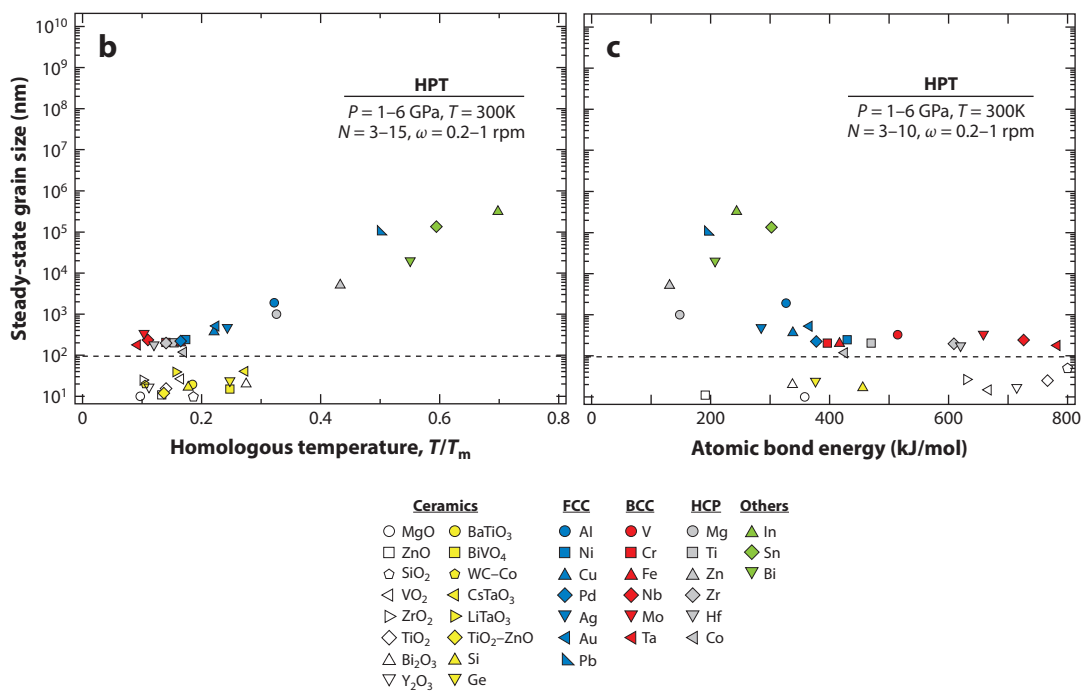

**Figure 7**

(*a*) High-resolution transmission electron microscopy image of ZnO (with starting structure of wurtzite) processed by HPT at room temperature under 6 GPa for 4 turns. Panel adapted with permission from Reference 136; copyright 2024 Elsevier. (*b*,*c*) Variations of steady-state grain size versus (*b*) homologous temperature and (*c*) atomic bond energy for HPT-processed ceramics compared with pure metals. Sources of data in panels *b* and *c* are given in **Table 1**, and the sources for pure metals are tabulated in References 161, 168, and 169. Abbreviations: BCC, body-centered cubic; FCC, face-centered cubic; HCP, hexagonal close-packed; HPT, high-pressure torsion.

zirconia compared with its monoclinic phase. Lattice defects can influence both the mechanical and functional properties of ceramics. A recent study suggested that even the inherent brittleness of ceramics can be significantly reduced by generating dislocations through an external source (8). Vacancies, dislocations, and grain boundaries can positively impact functional properties, particularly by altering the electronic structure of ceramics. Additionally, grain boundaries and grain

**Table 1  Major ceramics processed by HPT and their melting point ($T_m$), atomic (dissociative) bond energy, investigated features, number of HPT turns ($N$), grain size ($d$), and bandgap before and after HPT processing**

| Material | $T_m$ (K) (162, 163) | Bond energy (KJ/mol) (164) | Investigated features | $N$ | $d$ (nm) | Bandgap (eV) Before | After | Reference(s) |
|---|---|---|---|---|---|---|---|---|
| $Al_2O_3$ | 2,319 | 501.9 | Phase transition | 4 | 10–200 | 5.7 | 2.5 | 48, 139 |
| $BaTiO_3$ | 1,625 | ND | Dielectric | 5 | 20 | ND | ND | 74 |
| $Bi_2O_3$ | 1,090 | 337.2 | Photocurrent | 1 | 20 | 2.5 | 2.3 | 112 |
| $BiVO_4$ | 1,213 | ND | Photocatalysis | 4 | 15 | 2.4 | 2.3 | 85 |
| $CsTaO_3$ | 1,098 | ND | Photocatalysis | 1 | 40 | 4.6 | 3.3–3.7 | 137 |
| GaN–ZnO | 2,510.5[a] | ND | Photocatalysis | 3 | ND | 2.7 | 2.4 | 81 |
| $LiTaO_3$ | 1,923 | ND | Photocatalysis | 1 | 40 | 4.7 | 3–4.2 | 137 |
| MgO | 3,073 | 358.2 | Photocatalysis | 4 | >10 | ∼5.5 | 3.9 | 140 |
| $SiO_2$ | 1,610 | 799.6 | Photocatalysis | 5 | 10 | 9 | 2.8 | 141, 142 |
| $TiO_2$ anatase | 2,128 | 666.5 | Phase transition | 4 | 15 | ND | ND | 111 |
| $TiO_2$ anatase | 2,128 | 666.5 | Phase transition | 3 | 31 | ND | ND | 165 |
| $TiO_2$ anatase | 2,128 | 666.5 | Photocatalysis | 4 | ND | 3.2 | 2.4 | 75 |
| $TiO_2$ brookite | 1,843 | ND | Photocatalysis | 1 | ND | 3.1 | 2.8 | 134 |
| $TiO_2$ columbite | ND | ND | Photocatalysis | 3 | 32 | 3.2 | 2.5 | 135 |
| $TiO_2$ columbite | ND | ND | Photocurrent | 3 | 5–50 | 3.1 | 2.4 | 166 |
| $TiO_2$ columbite | ND | ND | Photocatalysis | 15 | ND | 3 | 2.4 | 160 |
| $TiO_2$–ZnO | 2,188[a] | ND | Photocatalysis | 15 | 12 | 3.2–3.3 | 1.6 | 80 |
| $VO_2$ | 1,818 | 629.7 | Phase transition | 1 | 27 | ND | ND | 141 |
| WC | 2,870 | ND | Consolidation | 10 | 20 | ND | ND | 167 |
| $Y_2O_3$ | 2,683 | 714.1 | Photoluminescence | 5 | 17 | 5.82 | 5.69 | 73 |
| ZnO | 2,248 | 191.5 | Photocatalysis | 3 | 11 | 3.1 | 1.8 | 76 |
| ZnO | 2,248 | ND | Photocatalysis | 4 | 9 | 3.1 | 2.9 | 136 |
| $ZrO_2$ | 2,963 | 766.1 | Phase transition | 10 | 25 | ND | ND | 47 |
| $ZrO_2$ | 2,963 | 766.1 | Photocatalysis | 2 | 27 | 5.1 | 4.3 | 138 |

Abbreviation: HPT, high-pressure torsion; ND, no data.

[a]The melting point of the composites was calculated as the average of the two components.

size play a crucial role in mechanical properties, as fine-grained ceramics typically exhibit higher strength and toughness. In this section, properties and applications reported for severely deformed ceramics are reviewed.

## 6.1. Consolidation of Powders

The application of HPT results in the consolidation of powders due to the high strain and pressure (5). While this produces bulk discs with high strength from metallic materials, the consolidation is only partial in ceramics due to their high hardness and poor ductility (46). In other words, HPT transforms ceramic powders into new powders with larger particle sizes and smaller surface areas. **Figure 8a,b** compares the appearance of $TiO_2$ powders before and after HPT processing, confirming that nanopowders were transformed into micropowders through partial consolidation (160). This partial consolidation has been utilized in some studies to reduce the sintering temperature and, consequently, increase the hardness of ceramics (68). **Figure 8d** compares the hardness of a ceramic-based composite, WC-11Co (wt%), without ($N = 0$) and with ($N = 10$) preconsolidation by HPT (167). The hardness levels achieved through preconsolidation by HPT are higher than those achieved without HPT processing and even exceed the hardness of the sample produced

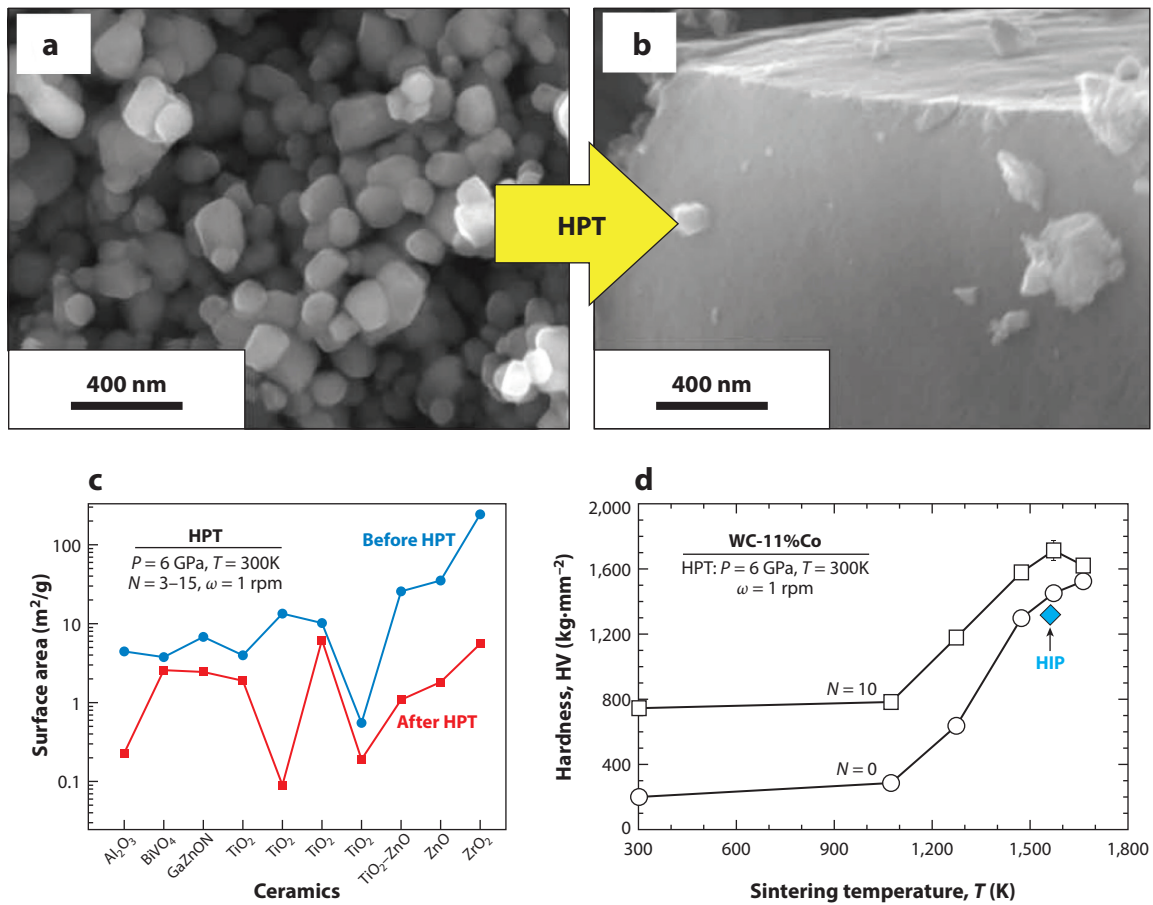

**Figure 8**

(*a,b*) Partial consolidation of TiO$_2$ by high-pressure torsion (HPT) processing at room temperature, examined by scanning electron microscopy (*a*) before and (*b*) after HPT processing. (*c*) Comparison of the specific surface area in Al$_2$O$_3$ (139), BiVO$_4$ (85), GaZnON (81), TiO$_2$ (75, 134, 135, 160), TiO$_2$–ZnO (80), ZnO (76), and ZrO$_2$ (138) before and after HPT processing. (*d*) Hardness of ceramic-based WC-11%Co (wt%) composite without ($N = 0$) and with ($N = 10$) preconsolidation by HPT before and after sintering at different temperatures compared with the hardness of the sample produced by hot isostatic pressing (HIP) at 1,663 K. Panels *a* and *b* adapted with permission from Reference 160. Panel *d* adapted from Reference 167.

by hot isostatic pressing at a similar sintering temperature (167). However, the consolidation of powders is a drawback when HPT-processed ceramics are used as catalysts, as the activity of catalysts is usually proportional to their specific surface area (75). As shown in **Figure 8c**, the specific surface area of many ceramics decreases after HPT processing (46). To overcome this drawback, a recent study attempted to enhance the surface area of HPT-processed ceramics through laser fragmentation, thereby improving their catalytic activity (171).

## 6.2. Photoluminescence

HPT processing can alter the optical properties of ceramics, including their photoluminescence (i.e., radiative recombination of photoexcited electrons and holes). The most common effect of HPT is reduction of the photoluminescence intensity, as shown in **Figure 9a** for TiO$_2$ after HPT processing (160). Such a reduction is associated with the introduction of defects, particularly

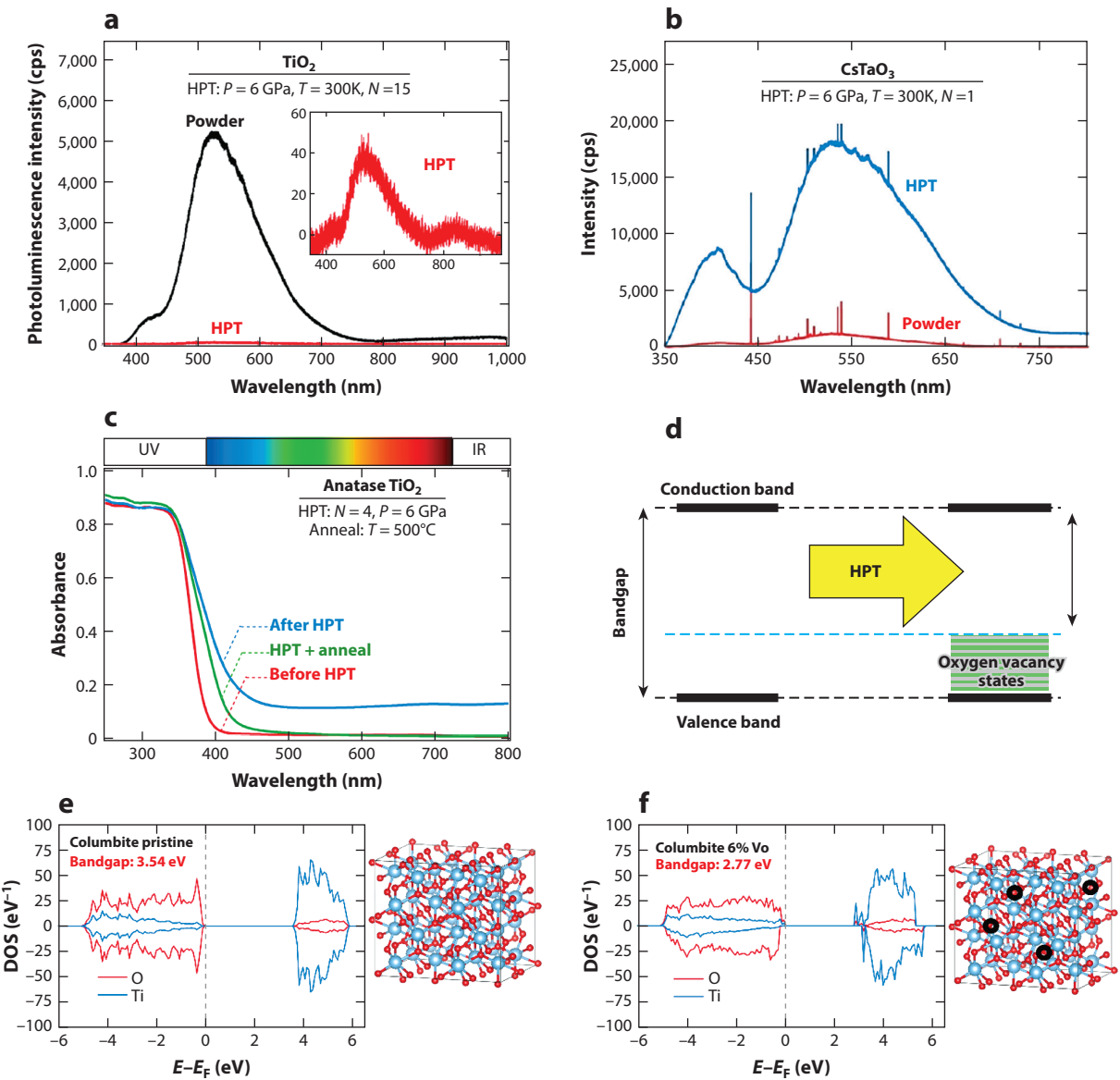

**Figure 9**

(*a,b*) Photoluminescence (*a*) reduction in TiO$_2$ (160) and (*b*) enhancement in CsTaO$_3$ (137) by high-pressure torsion (HPT) processing. (*c*) Enhancement of light absorbance in TiO$_2$ by HPT processing and its decrease after post-HPT annealing (75). (*d*) Mechanism of optical bandgap narrowing of oxide ceramics by HPT-induced oxygen vacancy generation. (*e,f*) Density functional theory calculation of the band structure of the high-pressure columbite phase of TiO$_2$ (*e*) without and (*f*) with oxygen vacancies, indicating a bandgap narrowing from 3.54 eV to 2.77 eV with 6% vacancy generation (vacancies are shown as *black circles* in supercells) (160). Panels *a*, *e*, and *f* adapted from Reference 160. Panel *b* adapted with permission from Reference 137. Panel *c* adapted with permission from Reference 75.

oxygen vacancies that act as shallow traps for the electrons, avoiding the radiative recombination (135, 160). This reduction of photoluminescence intensity was observed in other materials such as ZnO (136), BiVO$_4$ (85), and GaN–ZnO (81), which is a positive feature when ceramics are used as photocatalysts. However, some materials such as SiO$_2$ (142), CsTaO$_3$ (**Figure 9b**) (137), LiTaO$_3$ (137), BaTiO$_3$ (74), ZrO$_2$ (138), and Al$_2$O$_3$ (139) show an increase in photoluminescence intensity, with the enhancement becoming more significant as the number of HPT turns increases. For these oxides, the oxygen vacancies act as recombination sites, although this recombination is not necessarily detrimental to the photocatalytic activity of such oxides (137–139, 142).

## 6.3. Light Absorbance and Bandgap Narrowing

HPT processing of ceramics can enhance light absorbance, make their color darker (sometimes black), and reduce their bandgap, as summarized in **Table 1**. The introduction of vacancy point defects is the most important parameter influencing such optical property changes. As a striking example, it was shown that the bandgap of Al$_2$O$_3$, which is a wide-bandgap insulator, is reduced to 2.5 eV after HPT processing, and the oxide exhibits semiconducting properties in good agreement with the predictions of density functional theory (DFT) calculations (139). As schematically shown in **Figure 9d**, this bandgap narrowing is due to the formation of vacancy defect states between the conduction band and the valence band.

The combination of the phase transformation and large density of defects leads to further alterations in the light absorbance and bandgap of HPT-processed ceramics. As an example, anatase TiO$_2$, a well-known photocatalyst, can absorb only UV light due to its bandgap of 3.2 eV. The HPT method induces the high-pressure columbite phase and oxygen vacancies, resulting in narrowing of the bandgap to 2.7 eV as well as high light absorbance in the visible light region, as shown in **Figure 9c** (75). Calculations by DFT demonstrated the effect of oxygen vacancies on the bandgap narrowing of the columbite polymorph, as shown in **Figure 9e,f** (160). The increase in oxygen vacancy concentration in the bulk of these high-pressure polymorphs promotes the generation of shallow states near the conduction band [due to Ti$^{3+}$ formation (172, 173)] and deep defect states. The presence of vacancies and Ti$^{3+}$ narrows the optical bandgap of the columbite phase from 3.54 eV to 2.77 eV; however, bandgap narrowing is less significant when vacancies are generated in the anatase phase (from 3.09 eV to 2.93 eV with 6% oxygen vacancy generation) (160). Similar DFT results were achieved for higher bandgap narrowing of the high-pressure rock-salt phase of ZnO in the presence of vacancies (76). Further details about the simultaneous effect of vacancies and high-pressure phases on the bandgap of ceramics can be found elsewhere (72, 174, 175).

## 6.4. Photovoltaics and Photocurrent

One significant application of semiconductor ceramics in the clean energy sector is their use in solar cells or photoelectrochemical cells, owing to their photovoltaic and photocurrent performance (176, 177). TiO$_2$, as an n-type semiconductor, has been extensively researched for photovoltaic and solar cell applications. However, its primary drawback compared with silicon is its large bandgap, which restricts its activity to the UV region (178). To overcome this limitation for photovoltaic applications, the HPT method was employed to synthesize a vacancy-rich high-pressure columbite phase, known for its low bandgap and shifted Fermi level (166). The HPT-stabilized TiO$_2$-II phase generated photocurrent under visible light, with further enhancement observed after thermal annealing to optimize oxygen vacancy concentration, as shown in **Figure 10a**. In another study, black Bi$_2$O$_3$, rich in oxygen vacancies, was synthesized as a p-type semiconductor using HPT processing (112). As depicted in **Figure 10b**, photocurrent under visible light was achieved in HPT-processed Bi$_2$O$_3$, with further enhancement upon increasing the HPT processing temperature, due to an increase in oxygen vacancy concentration. While chemical methods and doping are commonly used to achieve visible-light photocurrent and photovoltaics, the HPT method provides a mechanical, dopant-free approach.

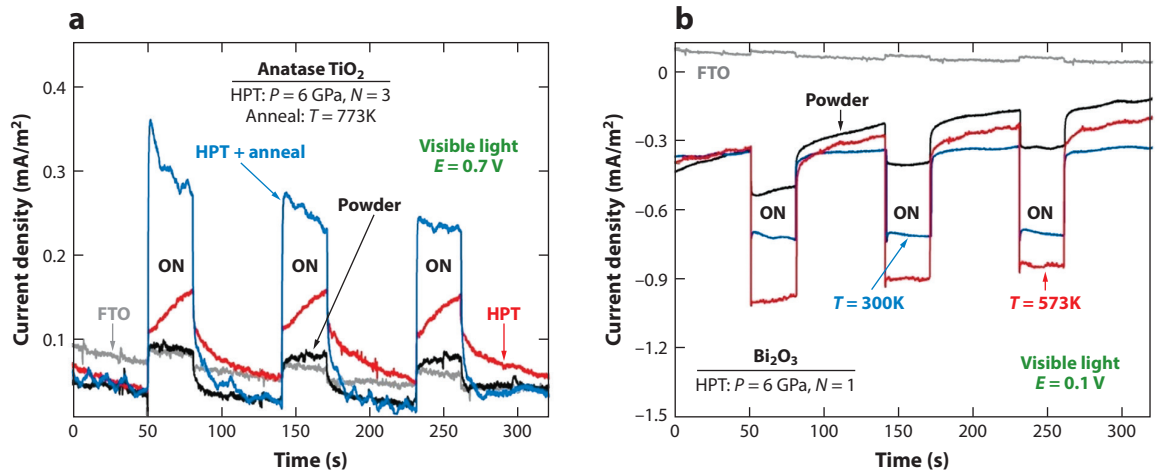

**Figure 10**

(*a*) Visible-light photocurrent in an n-type TiO$_2$ semiconductor ceramic processed by high-pressure torsion (HPT) with intensity enhancement after partial removal of oxygen vacancies by post-HPT annealing (166). (*b*) Visible-light photocurrent in a p-type Bi$_2$O$_3$ semiconductor ceramic by HPT processing with intensity enhancement by increasing HPT processing temperature and formation of black oxide (112). Fluorine-doped tin oxide glass (FTO) was used as substrate to make thin films of HPT-processed oxides. "ON" indicates switching on the light source. Figure adapted with permission from Reference 166.

## 6.5. Photocatalysis

Photocatalysis is a promising approach to addressing energy shortages and environmental pollution issues (176). Photocatalytic reactions occur in the presence of a catalyst and light, following four steps: (*a*) electron-hole generation from light absorption, (*b*) charge carrier separation, (*c*) charge carrier transfer to the catalyst surface, and (*d*) redox reactions on the surface. Photocatalysis can be used for various applications, including water splitting (180), CO$_2$ conversion (181), and hazardous substance degradation (182). The key challenge lies in developing photocatalysts with low bandgap electronic structures, extensive light absorption, low electron-hole recombination, and easy charge separation and migration. The HPT method has shown effectiveness in addressing many of these aspects through phase and defect engineering (183, 184). This section reviews some applications of HPT in photocatalysis.

**6.5.1. Toxic dye degradation.** HPT has been used to synthesize vacancy-rich Al$_2$O$_3$ ceramics for photocatalytic rhodamine B (RhB) dye degradation (139). **Figure 11***a* shows that the degradation efficiency improves with increasing HPT turns. HPT was also used to produce black Bi$_2$O$_3$ (112), oxygen vacancy-rich MgO for photocatalytic methylene blue dye degradation (140), and a high-pressure ZnO phase (76) for photocatalytic RhB dye degradation. Additionally, HPT-processed SiO$_2$ containing a high-pressure coesite phase, high-temperature amorphous phase, and oxygen vacancies exhibited enhanced dye degradation (142).

**6.5.2. Antibiotic degradation.** An HPT-synthesized CdS–TiO$_2$ composite catalyst, having sulfur vacancies and CdS/TiO$_2$ heterojunctions, showed high activity under visible light for the removal of tetracycline antibiotics, as shown in **Figure 11***b* (T.T. Nguyen & K. Edalati, manuscript in review).

**6.5.3. Hydrogen production.** HPT-processed Ga$_6$ZnON$_6$ oxynitride [shown in **Figure 11***c* (81)], LiTaO$_3$ (137), and CsTaO$_3$ (137) were reported to have good photocatalytic hydrogen production performance due to the presence of vacancies. Another group of HPT-produced

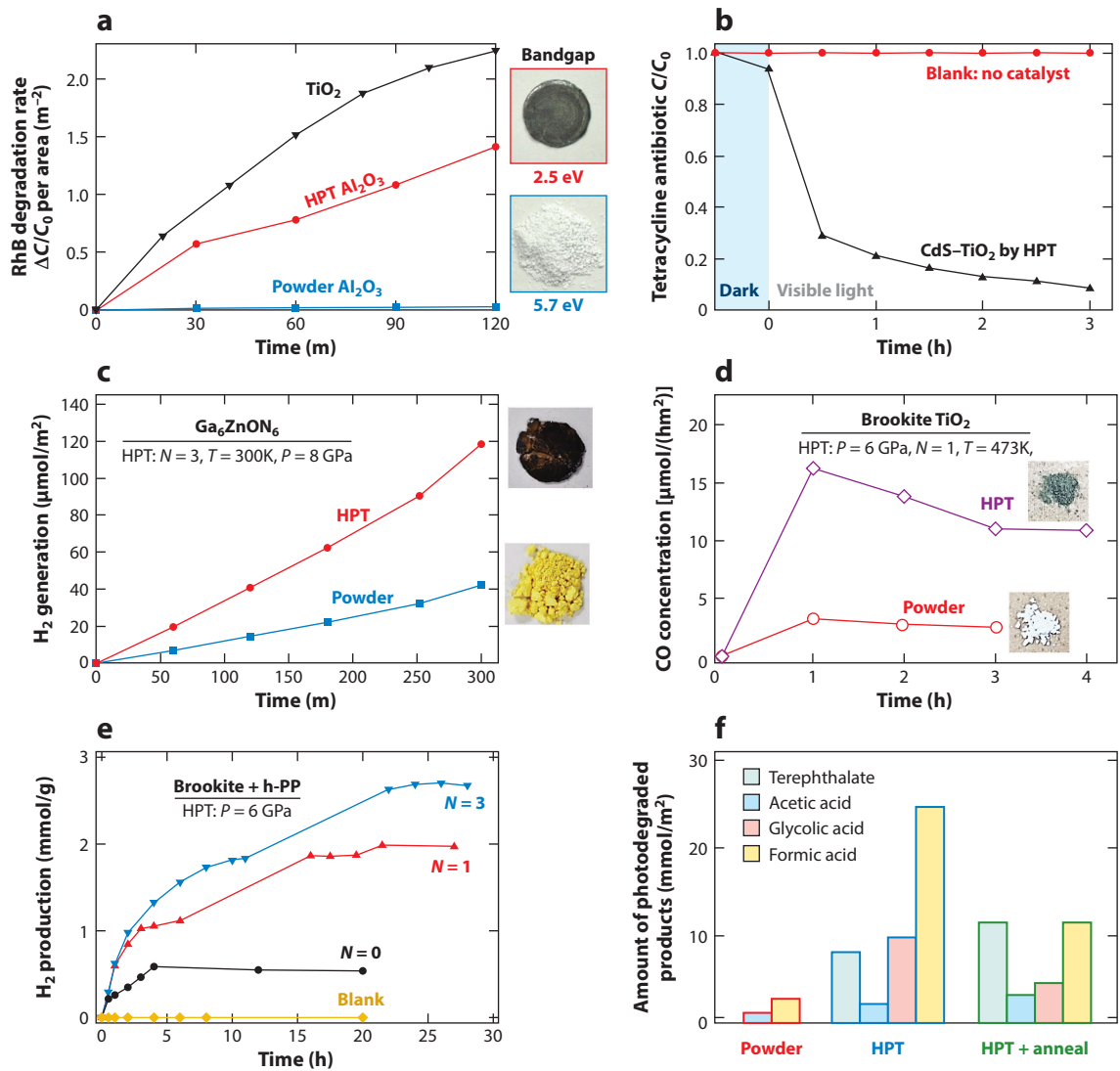

**Figure 11**

Photocatalytic activity of ceramics processed by high-pressure torsion (HPT). (*a*) Rhodamine B (RhB) dye degradation on HPT-processed Al$_2$O$_3$ in comparison with initial Al$_2$O$_3$ powder (no activity) and TiO$_2$. Panel adapted with permission from Reference 139; copyright 2019 Acta Materialia. (*b*) Tetracycline antibiotic degradation under visible light using an HPT-synthesized CdS–TiO$_2$ composite (T.T. Nguyen & K. Edalati, manuscript in review). (*c*) Hydrogen production from water splitting using HPT-processed Ga$_6$ZnON$_6$ oxynitride in comparison with initial powder. Panel adapted with permission from Reference 81; copyright 2019 Acta Materialia. (*d*) CO production from CO$_2$ conversion on HPT-processed black brookite compared with initial powder. Panel adapted with permission from Reference 134; copyright 2023 Elsevier. (*e*) Hydrogen production from photoreforming of polypropylene (h-PP) plastic on brookite TiO$_2$ without ($N = 0$) and with ($N = 1, 3$) HPT processing for different turns. Panel adapted with permission from Reference 185; copyright 2024 Hydrogen Energy Publications LLC. (*f*) Degradation of polyethylene terephthalate plastic to small organic molecules (formic acid, terephthalate, glycolic acid, and acetic acid) by photoreforming using TiO$_2$ without (powder) and with HPT processing (HPT) and after post-HPT annealing at 773 K for 1 h. Panel adapted from Reference 186; copyright 2024 Elsevier. Images in panels *a*, *c*, and *d* show the appearance of samples before and after HPT processing, and $C_0$, $C$, and $\Delta C$ in panels *a* and *b* indicate initial concentration, final concentration, and degraded amount of dye or antibiotic, respectively. The disc samples in panels *a* and *c* were crushed to the powder form before the photocatalytic test.

photocatalysts with high photocatalytic hydrogen activity includes metastable or high-pressure phases of $TiO_2$ (75, 80, 160), $ZrO_2$ (138), and ZnO (136). Moreover, high-entropy ceramics processed by HPT, including $TiZrHfNbTaO_{11}$ and $TiZrHfNbTaO_6N_3$, also possess superior photocatalytic hydrogen production (82, 84, 171, 187), while $TiZrNbTaWO_{12}$ with multiple heterojunctions is more appropriate for oxygen production (188).

**6.5.4. $CO_2$ conversion.** Various photocatalysts produced by HPT for $CO_2$ conversion include $BiVO_4$ (85), black brookite $TiO_2$ (134), high-pressure columbite $TiO_2$ (135), and high-entropy ceramics (83, 189). The black brookite $TiO_2$, synthesized by HPT at a temperature of 473 K, exhibits abundant oxygen vacancies and high performance for $CO_2$ to CO conversion, as shown in **Figure 11$d$** (134). The activity of black brookite was even higher than P25 $TiO_2$, which is considered a benchmark in the photocatalysis field.

**6.5.5. Photoreforming of plastic wastes.** Photoreforming is an important process that converts plastic waste into value-added organic molecules while simultaneously producing hydrogen fuel. Since brookite $TiO_2$ was reported to be an active photocatalyst for the photoreforming of polyethylene terephthalate (PET) plastic (190), a recent study used brookite for the photoreforming of nonrecyclable polypropylene plastic waste (185). The study showed that hydrogen production from photoreforming increases with increasing HPT turns from 1 to 3, accompanied by the formation of small organic molecules, as shown in **Figure 11$e$**. The high activity was due to the generation of strain-induced defects in both the photocatalyst and plastic waste, as well as the improvement of charge carrier mobility between the catalyst and plastic. Another study showed that HPT-processed $TiO_2$, which contains oxygen vacancy–rich anatase and columbite phases, exhibits higher activity compared with unprocessed anatase powder for the photoreforming of PET plastic into valuable organic compounds such as formic acid, terephthalate, glycolic acid, and acetic acid (186). However, as shown in **Figure 11$f$**, its activity decreases with annealing at 773 K for 1 h due to the annihilation of oxygen vacancies (186).

## 6.6. Electrocatalysis

Efficient catalytic materials for the oxygen evolution reaction, oxygen reduction reaction, and hydrogen evolution reaction are crucial for sustainable electrochemical energy applications, particularly through electrochemical water splitting (191, 192). Precious metals, their alloys, and their oxides are typically used for electrocatalytic water splitting (193, 194). Recent studies have focused on cost-effective and environmentally friendly catalysts such as $TiO_2$, attempting to enhance its electrocatalytic activity through doping (195) or the addition of metallic particles (196). Building on the success of enhancing the photocatalytic activity of $TiO_2$ HPT processing (75), another study demonstrated the effectiveness of HPT in enhancing the electrocatalytic properties of pure $TiO_2$ (197). As shown in **Figure 12$a$**, the inclusion of a ~20 wt% high-pressure columbite phase in $TiO_2$ via HPT processing can enhance electrocatalytic activity to a maximum level not only for hydrogen production from water but also for the hydrogenation of oxalic acid to produce an alcoholic compound (197). This improvement was attributed to the columbite's crystal structure and electronic band structure as well as its contribution to forming heterojunctions in the anatase matrix. The enhancement of electrocatalytic activity through HPT processing was also reported in Fe–Si–B glasses for both hydrogen evolution (198) and oxygen evolution (199) reactions.

## 6.7. Thermoelectric Properties

For many years, ceramics and oxides have been discussed for their use as thermoelectric materials (200, 201). So far, the HPT method has been mainly applied to skutterudites, half-Heusler alloys,

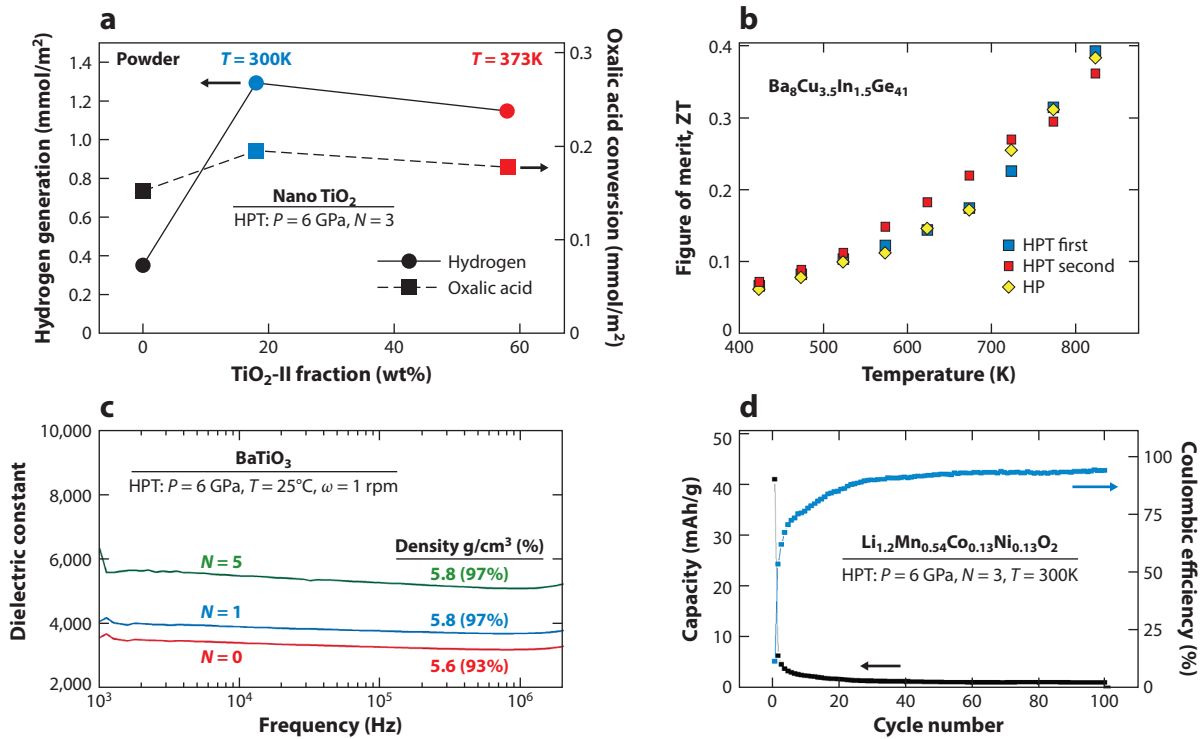

**Figure 12**

(*a*) Enhancement of electrocatalytic activity of $TiO_2$ for hydrogen formation and oxalic acid conversion by formation of the $TiO_2$-II phase through high-pressure torsion (HPT), particularly after processing at 300 K. Panel adapted from Reference 197 (CC BY 4.0). (*b*) Thermoelectric figure of merit ZT versus temperature for a $Ba_8Cu_{3.5}In_{1.5}Ge_{41}$ clathrate ceramic after HPT processing compared with hot pressing (HP), where "HPT first" and "HPT second" refer to warming after HPT and cooling following the warming after HPT, respectively (205). (*c*) Variation of dielectric constant versus frequency at ambient temperature for a $BaTiO_3$ ceramic processed by HPT for 0, 1, and 5 turns and sintered at 1,473 K for 2 h (with density and relative density of samples given). Panel adapted from Reference 74 (CC BY 4.0). (*d*) Cycle performance of a $Li_{1.2}Mn_{0.54}Co_{0.13}Ni_{0.13}O_2$ as a cathode material of Li-ion batteries (206).

and Bi tellurides to enhance their figure of merit (ZT) because of their high original thermoelectric efficiency (202). In fact, record ZT values of ~2.1 for skutterudites (203), ~1.9 for Bi tellurides (202), and ~1.5 for half-Heusler alloys (204) could be reached by HPT processing (ZT $= S^2 T/\rho\kappa$, where $S$, $\rho$, $T$, and $\kappa$ are the Seebeck coefficient, electrical resistivity, absolute temperature, and thermal conductivity, respectively). These values were achieved because of the significant decrease in the thermal conductivity, which overcompensated for the increase of electrical resistivity, with both changes arising from the enhanced density of SPD-induced lattice defects.

In contrast to the ceramic-like materials mentioned above, ceramics such as clathrates (205) and perovskites (207) exhibit a comparably low initial thermal conductivity that limits the potential for its further decrease and thus for an increase of ZT. In the case of clathrates, the electrical resistivity resulting from SPD-induced defects (i.e., not only from grain boundaries but also from free dislocations in particular) seems to outweigh the ZT benefits from increases of thermopower and decreases of thermal conductivity. When the density of free dislocations is reduced in repeated heating cycles to 670 K (homologous temperature of ~0.6 $T_m$), the ZT value of HPT-processed clathrate is increased by about 35% at 670 K, as shown in **Figure 12b**. This phenomenon has also been observed in the (half-)Heusler alloys (208).

Another example of promising modifications of thermoelectric properties of ceramics by means of HPT has been recently presented for a lead-free halide perovskite (a candidate for printed electronics because of high charge carrier mobility, long diffusion length of charge carriers, and ultralow thermal conductivity) (207). The potential of SPD-mediated improvement of thermoelectric performance in halide perovskites looks somewhat higher than that of clathrates as the number of SPD-induced dislocations is limited. This fact and the marked increase in charge carrier mobility through SPD reduced the electrical resistivity by ~40% compared with the unprocessed sample, thus extending the thermopower by ~30% relative to the starting material.

## 6.8. Dielectric Properties

Dielectric properties refer to the ability of a material to store and dissipate electrical energy within an electric field, characterized by its dielectric constant and loss factor (209). These properties enable the efficient functioning of capacitors in electronic devices, such as smartphones and computers (209). It was shown that HPT processing can enhance the dielectric constant of $BaTiO_3$, with the enhancement becoming more significant as the number of HPT turns increases, as illustrated in **Figure 12c** (73). A dielectric constant of over 5,000 achieved in $BaTiO_3$ through HPT processing surpasses the values reported for conventional processes (210, 211). This increase after HPT is attributed to the effects of tetragonality, grain size, and resultant changes in the elastic field and domain wall energy (210). Moreover, powder precompaction by HPT processing (212) is expected to contribute to better dielectric properties. The increase in the dielectric constant for well-known $BaTiO_3$ dielectric materials is significant for developing high-capacitance multilayered ceramic capacitors (210).

## 6.9. Other Properties and Applications

Several other studies have investigated the properties and potential applications of oxides after HPT processing, including microstructure and phase stability (213, 214), magnetic properties (215), ferroelectric properties (216, 217), heat capacity (218), optical properties (219, 220), electronic band structure features (221), and radiation absorbance (222). These studies indicate that HPT processing can modify the structural and microstructural features of ceramics to achieve desirable functional properties. There are also reports on the application of HPT-processed ceramics as anode (223) and cathode (206) materials for Li-ion batteries. An example is shown in **Figure 12d**, indicating that the initial specific discharge capacity for the cathode material $Li_{1.2}Mn_{0.54}Co_{0.13}Ni_{0.13}O_2$ processed by HPT is as low as 40 mAh/g, and the coulombic efficiency does not reach 100% after 100 cycles (206). These negative results are due to powder consolidation by HPT processing, whereas nanopowders are needed to make effective cathodes. Despite these preliminary negative data, HPT-processed oxides have a high potential for application in Li-ion batteries due to the positive effect of strain-induced oxygen vacancies on the electrochemical performance of cathode materials (224).

# 7. SYNTHESIS OF NEW CERAMICS BY HIGH-PRESSURE TORSION

The HPT process is not only used as a processing tool for existing ceramics but also contributes to the synthesis of new ceramics, such as black oxides, metastable high-pressure ceramics, and high-entropy ceramics.

## 7.1. Black Oxides

Black oxides are oxides with large concentrations of oxygen vacancies and high light absorbance, which have become popular as photocatalysts (225, 226). These materials are usually synthesized by chemical methods such as annealing under a hydrogen atmosphere, but HPT provides a mechanical route to produce such vacancy-rich black oxides. Black oxides have been successfully produced by HPT from $Al_2O_3$ (139), $TiO_2$ (134), $ZrO_2$ (138), and $Bi_2O_3$ (112, 227). All these black oxides exhibit high light absorbance and enhanced photocatalytic activity.

## 7.2. Metastable Ceramics

High pressure and high strain during HPT processing can lead to phase transformations in ceramics (42, 62–64). The generation of lattice defects and the nanograin size effect after HPT processing can lead to the stabilization of high-pressure phases even after releasing pressure (46). Such metastable phases have been reported to exhibit interesting properties, such as photocatalytic activity (75, 76, 228), electrocatalytic activity (197), and dielectric properties (74). Moreover, HPT can lead to the discovery of new phases that are hidden under normal straining or compression conditions (45). Sheared $TiO_2$ phases are examples of such phases, which were theoretically predicted and experimentally detected as intermediate phases during HPT-induced phase transformations (113).

## 7.3. High-Entropy Ceramics

High-entropy ceramics are defined as ceramics containing at least five principal cations with a mixing entropy greater than 1.5R (R being the gas constant) (229, 230). A main feature of high-entropy photocatalysts is their high stability, which makes them promising for various applications (231). Various high-entropy ceramics, including high-entropy oxides (83, 188) and high-entropy oxynitrides (84, 189), have been discovered by the HPT method for photocatalytic applications. In fact, the first high-entropy photocatalyst reported in the literature was synthesized by HPT (82). High-entropy oxides were synthesized by a two-step process (83, 188): (*a*) the application of ultra-SPD [i.e., shear strains over 1,000 (232)] to a metallic powder mixture for mechanical alloying and (*b*) high-temperature heating for oxidation. For high-entropy oxynitride, a third step of high-temperature nitriding should be applied (84, 189). These materials were used for hydrogen production (82, 84, 171, 233, 234), oxygen production (188), $CO_2$ conversion (83, 189, 233, 234), and plastic degradation (235, 236), although a wider range of applications is expected in the future (229, 230).

## 8. CONCLUDING REMARKS AND OUTLOOK

This review highlights the transformative impact of SPD on ceramic materials, focusing on phase and microstructural evolution as well as the enhancement of their mechanical, electrical, and optical properties. The success in SPD processing of ceramics by HPT underscores the potential of the technique for producing high-performance ceramics suitable for various applications, from structural components to advanced energy materials. Despite significant progress in SPD processing of metallic materials (Section 1) and high-temperature deformation of ceramics (Section 2), SPD processing of ceramics at ambient temperature is a rather less-known topic in the materials science field (Section 3). SPD can induce unique phase transformations (Section 4) and microstructural features such as high vacancy and dislocation concentrations and nanograins (Section 5), which are barely achievable through conventional processing methods. These microstructural modifications result in ceramics that exhibit enhanced consolidation and mechanical properties as well as improved functional properties, including better photocatalytic, electrocatalytic, thermoelectric, and dielectric performance (Section 6). The introduction of novel black, high-pressure, and high-entropy ceramics by SPD further expands the potential applications of these materials in harsh environmental conditions and energy sectors (Section 7).

Several key areas of research can advance the field of SPD-processed ceramics. First, integrating advanced characterization techniques, such as in situ transmission electron microscopy and synchrotron XRD, can provide deeper insights into the real-time microstructural evolution and phase transformations occurring during the SPD processing of ceramics. Such studies will enable more precise control over processing parameters to achieve the desired microstructures and properties. Second, the exploration of new ceramics, such as high-entropy ceramics with tailored compositions, can be pursued. Due to their multicomponent nature, these materials offer a vast compositional space for optimizing properties and performance. Third, the application of theoretical calculations, machine learning, and artificial intelligence in the design and optimization of SPD-processed ceramics can help identify potential ceramics that may not be apparent through traditional analyses. Lastly, expanding the applications of SPD-processed ceramics beyond traditional domains (e.g., the development of ceramics for biomedical applications) and into environmental applications (e.g., water purification, gas separation, and microplastic cycling) can lead to significant societal benefits. Continued interdisciplinary research combining materials science, chemistry, and computational techniques can facilitate the development of the next generation of SPD-processed ceramic materials.

## DISCLOSURE STATEMENT

The authors are not aware of any affiliations, memberships, funding, or financial holdings that might be perceived as affecting the objectivity of this review.

## ACKNOWLEDGMENTS

K.E. thanks the Japan Society for the Promotion of Science for a Grant-in-Aid (JP22K18737). J.H.-J. acknowledges the Q-Energy Innovator Fellowship of Kyushu University, the Japan Science and Technology Agency, and the Establishment of University Fellowships Towards the Creation of Science Technology Innovation (JPMJFS2132). T.T.N. is supported by Mitsui Chemicals, Inc., Japan. N.E. acknowledges support from the Ministry of Science and Higher Education of the Russian Federation as part of the World-Class Research Center Program: Advanced Digital Technologies (contract 075-15-2022-313 dated April 20, 2022). V.I.L. appreciates support from the National Science Foundation (CMMI-1943710, DMR-2246991, and MSS170015), Army Research Office (W911NF2420145), and Iowa State University (Murray Harpole Chair in Engineering). R.Z.V. acknowledges support from the Russian Science Foundation (24-43-20015).